\documentclass[12pt]{article}
\usepackage{psfig}
\newcommand{\be}{\begin{equation}}
\newcommand{\ee}{\end{equation}}
\newcommand{\bea}{\begin{eqnarray}}
\newcommand{\eea}{\end{eqnarray}}
\begin{document}
\title{Application of time-dependent density functional theory to
electron-vibration coupling in benzene}
\author{
A. Schnell, G. F. Bertsch\thanks{E-mail: bertsch@phys.washington.edu}\\
{\it Institute for Nuclear Theory, University of Washington, Seattle, WA 98125, USA}\\[3mm]
K. Yabana\\
{\it Institute of Physics, University of Tsukuba, Tsukuba 305-8571, Japan}
}
\maketitle
\def\pipi{$\pi$-$\pi^*$}
\begin{abstract}
Optical properties of symmetry-forbidden \pipi~ transitions in benzene
are calculated with the time-dependent density functional theory (TDDFT),
using an adiabatic LDA functional.  Quantities calculated are the envelopes
of the Franck-Condon factors of the vibrationally promoted transitions and
the associated oscillator strengths. The strengths, which span three
orders of magnitude, are reproduced to better than a factor of two by
the theory.  Comparable agreement is found for the Franck-Condon widths.
We conclude that rather detailed information can be obtained with the
TDDFT and it may be worthwhile to explore other density functionals.

\end{abstract}

The time-dependent density functional theory (TDDFT) has proven to be
a surprisingly successful theory of excitations and particularly
the optical absorption strength function.  The theory is now
being widely applied in both chemistry and in condensed matter physics.
The literature in quantum chemistry is cited in a recent study on the 
electronic excitations in benzene \cite{he00}.  Benzene is an interesting
molecule for testing approximations because its spectra have been
very well characterized, both electronic and vibrational.  In this
note we will apply the TDDFT to coupling between vibrational and
electronic excitations. In our previous studies, we have investigated many different electronic
structure questions using a rather simple version of the density functional
theory, the local density approximation (LDA).  Our emphasis has been to study
the overall predictive power of a fixed functional rather than to try
to find the best functional for each properties.  The approximation scheme we 
consider is straightforward and uses the same computer programs as for
calculating purely electronic excitations.  We treat the electronic dynamics in the 
adiabatic approximation, taking the same energy function for the
dynamic equation as is used in the static structure calculation.  In
our view, this is the only consistent scheme available that guarantees
conservation of the oscillator sum rule.  The electron-vibration coupling is
treated in a vertical approximation, so only information at frozen nuclear
coordinates is required.

We consider only spin-singlet states in this work and drop the spin
designation in labeling the states. Empirically, the lowest 
states derive from the \pipi~ manifold, exciting an electron from
the
two-fold degenerate $e_{1g}$ HOMO orbital to the two-fold degenerate
$e_{2u}$ orbital.  The four states consist of a strongly absorbing
two-fold degenerate
$E_{1u}$ excitation and
two other states, $B_{1u}$ and $B_{2u}$, for which symmetry forbids any
transition strength.  This basic spectrum is shown in Fig. 1, comparing
also with our TDDFT calculation, the TDDFT calculation of ref. \cite{he00},
and the CI theory of ref. \cite{ma87}.

\begin{figure}[tb]
\begin{center}
\psfig{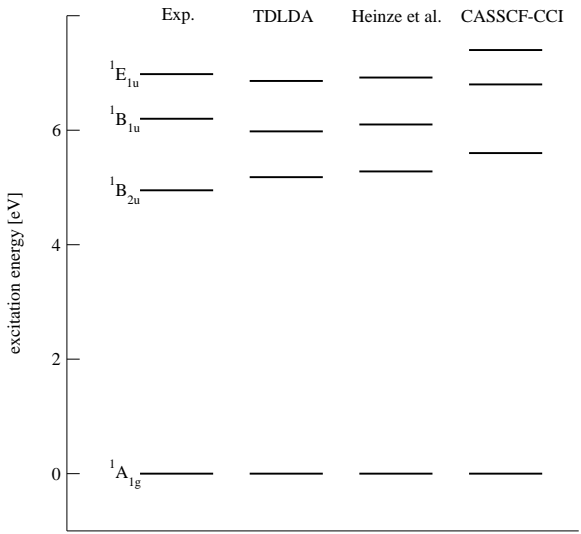}
\end{center}
	\caption{Electronic excitations of benzene in the
\pipi~ manifold.  Besides the experimental data and the
present TDDFT, we show the TDDFT of ref. \protect\cite{he00}
and the CI calculation of ref. \protect\cite{ma87}}
\end{figure}
It is seen that the TDDFT gives an
excellent account of the energies. In fact the TDDFT gives a good description of
the higher frequency absorption including $\sigma$-$\sigma^*$ transitions
as well \cite{YB1999}.  
The detailed optical properties of the three transitions have been
studied gas phase absorption \cite{pa78,hi91}. The strong transition is
the $E_{1u}$ with $f=0.9-0.95$. The $B_{1u}$ mode is seen as a shoulder on the strong $E_{1u}$ 
peak.  Its total transition strength is
about a factor of 10 lower than the strong state; ref. \cite{pa78}
quotes a value $f=0.09$. 
The
$B_{2u}$ transition is very weak and is seen as a partially resolved set of 
vibrational transitions with a total strength about
$f \approx 1.3 \times 10^{-3}$\cite{pa78}.  The strength associated with
the most prominent resolved states is $0.6 \times 10^{-3}$\cite{hi91}.

The vibrational couplings of the $B$ states has been recently studied using
the CASSCF method and analytic expressions for the linear coupling to
vibrations\cite{be00}, and we shall compare with their results.  The 
TDDFT includes correlation effects in a  different way, and has some well-known advantages such as the automatic
conservation of required sum rules.  Also, as mentioned earlier, the present
method does not require  any reprogramming.

For our treatment of the vibrational motion, we assume that 
the the vibrations are harmonic in the electronic ground state.  The 
Hamiltonian may be defined
\be
{\cal H} = -\sum_i^{3N} {\hbar^2\over 2 M_i m}{\partial^2\over\partial q_i^2}
+\frac{1}{2}\sum_{ij}^{3N} F_{ij}q_iq_j.
\ee
where $q_i$ are the 36 Cartesian 
displacement coordinates of the 12 atomic centers,
$m$ is the atomic mass unit, $M_i$ is the mass of the atom in 
daltons,
and $F_{ij}$ is the matrix of force constants.  
The matrix ${\bf M}^{-1/2}{\bf F} {\bf M}^{-1/2}$ ({\bf M} is the diagonal
matrix of masses $M_i$) is diagonalized by an orthogonal
transformation ${\bf U}$ to obtain
the normal modes $Q_k$ and the eigenfrequencies $\omega_k=2\pi \nu_k$.
The Cartesian displacements are obtained directly from the rows of the
transformation matrix ${\bf U}$, $q_i = M^{-1/2}_i\sum_kU_{ik} Q_k$.
The translational and rotational motions will also
be contained in the transformation matrix ${\bf U}$ as zero frequency modes.
The probability distribution of the zero point motion is then
given simply by
\be
P({\vec Q}) \sim \exp (-\sum_k Q_k^2/2 Q_{0k}^2).
\ee  
where $$Q_{0k} = \sqrt{ {\hbar \over 2 m \omega_k}} = {4.1 \hbox{[\AA]}\over
\sqrt{n_{cm}}}
$$ 
is
the r.m.s.~amplitude of the zero-point motion\footnote{At finite temperature
the r.m.s.~amplitude is increased by a factor $1/\sqrt{\tanh(\hbar \omega_k/
2 k_B T)}$.}. The last equality expresses the formula in common units with
$n_{cm}=c/\nu$ the energy of the vibration in wavenumbers [cm$^{-1}$].

The optical absorption strength function in the presence of the zero point
motion is determined by the convolution 
the probability distribution of displacements with the strength
calculated as a function of displacement,
\be\label{conv}
f = \int d^N Q_k  P({\vec Q }) f({\vec Q}).
\ee
We thus need the
absorption strength as a function of the normal mode coordinates
$Q_k$.  In the case of a forbidden transition promoted
by the vibration $k$, the coupling is linear for small displacements
and the transition
strength will be quadratic in  $Q_k$,
\be\label{f_0k}
f(Q_k)  = f_{0k} {Q_k^2 \over Q_{0k}^2}+...
\ee
We verify below that this functional dependence is satisfied for the
couplings of interest in benzene.  Then the
convolution over the
ground state probability distribution gives simply 
$f = f_{0k}$.

We also consider widths of the transitions due to the Franck-Condon 
factors of multiply excited vibrations.  This is calculated by replacing
$f$ with the strength function $S(E,Q) = f(Q) \delta(E-E(Q))$ in eq.
\ref{conv}. Assuming that the excitation energy is linear in $Q$,
$$
E(Q_k) \approx E_0 + K_k Q_k/Q_{0k}+...
$$
the Gaussian probability distribution $P$ gives a Gaussian envelope
for the Franck-Condon factors,
\be\label{P(E)}
P(E) \sim \exp ( - (E-E_0)^2/2 K_k^2).
\ee

\begin{figure}[tb]
\begin{center}
\psfig{figure=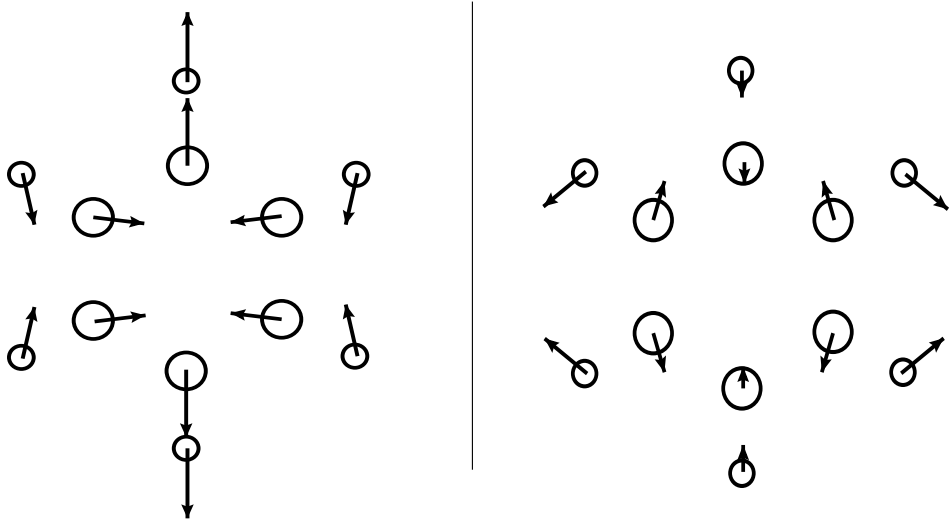,height=7cm}
\end{center}
\caption{Cartesian 
displacements of the vibrational modes 6a (left) and 8a (right).  
These modes have symmetry $E_{2g}$ and give the most important couplings for our
purposes.  The r.m.s.~displacement of the atoms are magnified by a factor 40, 
i.e.  $Q_k = 40 Q_{0k}$ with respect to the scale for the equilibrium 
positions.}
\label{fig_modes}
\end{figure}

For the numerical studies reported here, we constructed the
transformation matrix
${\bf U}$ using the empirical force field of Goodman and
Ozkabak \cite{Goodman}, which fits the observed frequencies extremely
well. {\it Ab initio} calculations of the force field have also
reached a high level of accuracy
\cite{MTL1997}. However, as mentioned earlier, we do not make our own
DFT calculations of the force constants because our goal is the dynamic
behavior of the electrons. The frequencies and symmetries of the normal
modes are listed in Table 1, taken from ref. \cite{Goodman}.  The most
important modes
for the induced transition strengths are the $E_{2g}$ and the
$B_{2g}$ modes\footnote{The $B_{1g}$ symmetry would also give couplings
between the electronic states, but there are no vibrations of that
symmetry.}. The $E_{2g}$ vibrations couple the strong electronic excitation
to the other states in the
\pipi~ manifold.  The $B_{1g}$ can induce out-of-plane dipole strength for
these excitations.  The theoretical widths of the excitations are largely due 
to mode 1, which is an $A_{1g}$ radial oscillation mode that favors  
carbon displacements. In 
Fig.~\ref{fig_modes} we show the Cartesian displacements associated
with the two strongest $E_{2g}$ modes with respect to carbon displacements.  

The present TDDFT calculations were performed making use of the
same representation of the Kohn-Sham operator as in our previous study
of the full energy distribution in optical absorption \cite{YB1999}. 
The wave functions are represented on a coordinate-space mesh as
has been introduced in condensed matter physics \cite{ch??}. 
However, the algorithm in the present program is a new one \cite{ya00} that uses the conjugate
gradient method to extract individual states rather that the direct
real-time propagation of the wave function.  While the real-time method
is very efficient for calculating the global strength function, it is less
suited for locating individual eigenstates when they are weakly excited
by the dipole operator.  
In both methods, the electronic ground state for a given nuclear geometry
is first computed with the Kohn-Sham equation,
$$
-{\nabla^2 \over 2 m}\phi_i +{\delta {\cal V}\over\delta n}\phi_i=
\epsilon_i\phi_i.
$$ 
We use a simple LDA  energy density
functional \cite{pe81} for the electron-electron interaction in ${\cal V}$ and a 
pseudopotential approximation \cite{tr91,kl82} to treat the interaction
of the valence electrons with the ions.  The important numerical 
parameters in the calculation are the  
the mesh spacing, taken as $\Delta x = 0.3 $ \AA, and the volume in which the
wave functions are calculated, which we take as a sphere of radius 7 \AA.
With these parameters, orbital energies are converged to better than 
0.05 eV.    
\begin{figure}[tb]
\begin{minipage}[t]{7.5cm}
\psfig{figure=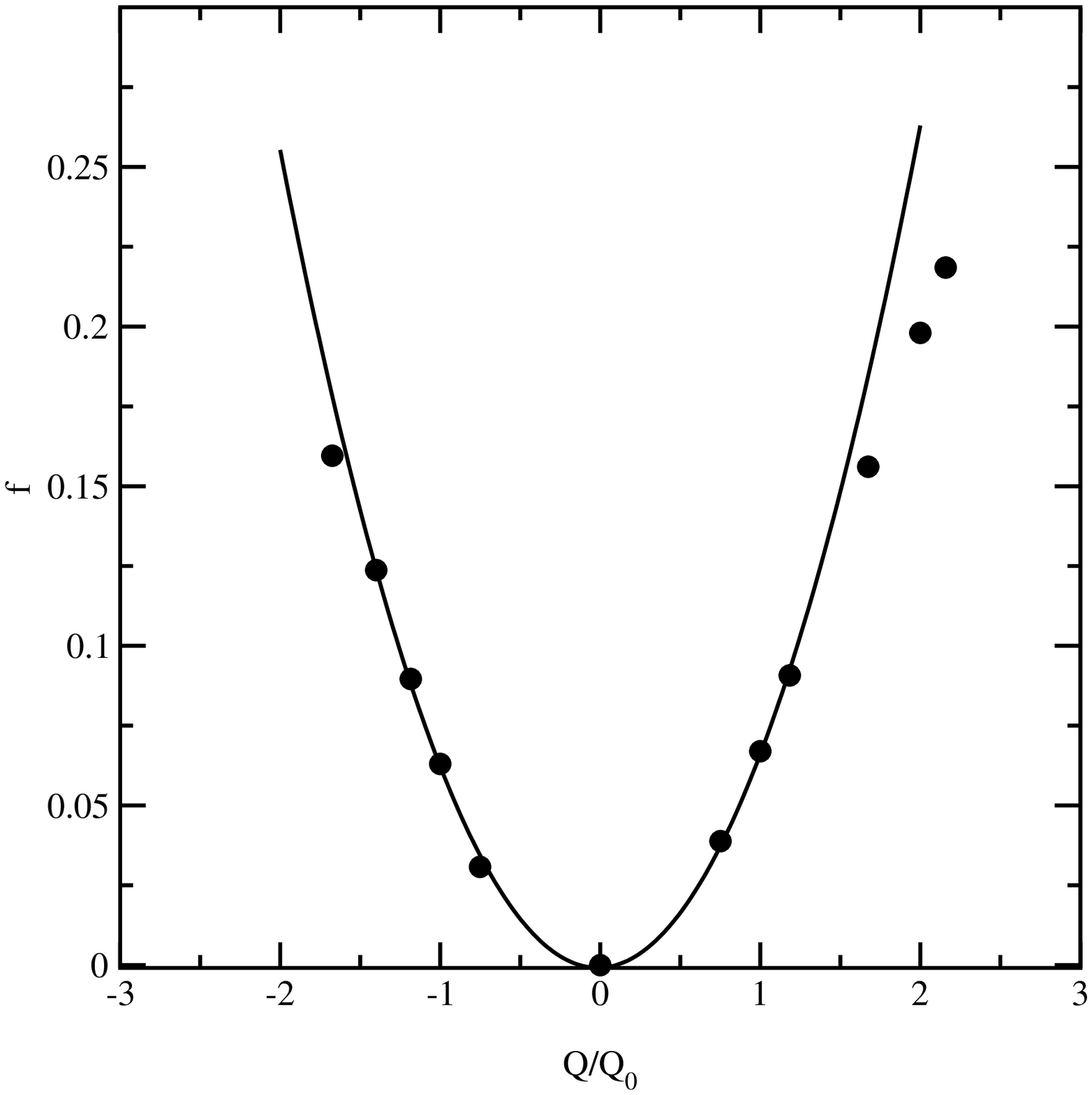,height=6.5cm}
\caption{Dependence of the oscillator strength of the $
^1B_{2u} \leftarrow ^1A_{1g}$ transition on the vibrational coordinate
for the 8a mode.}
\label{8a}
\end{minipage}
\hfill
\begin{minipage}[t]{7.5cm}
\psfig{figure=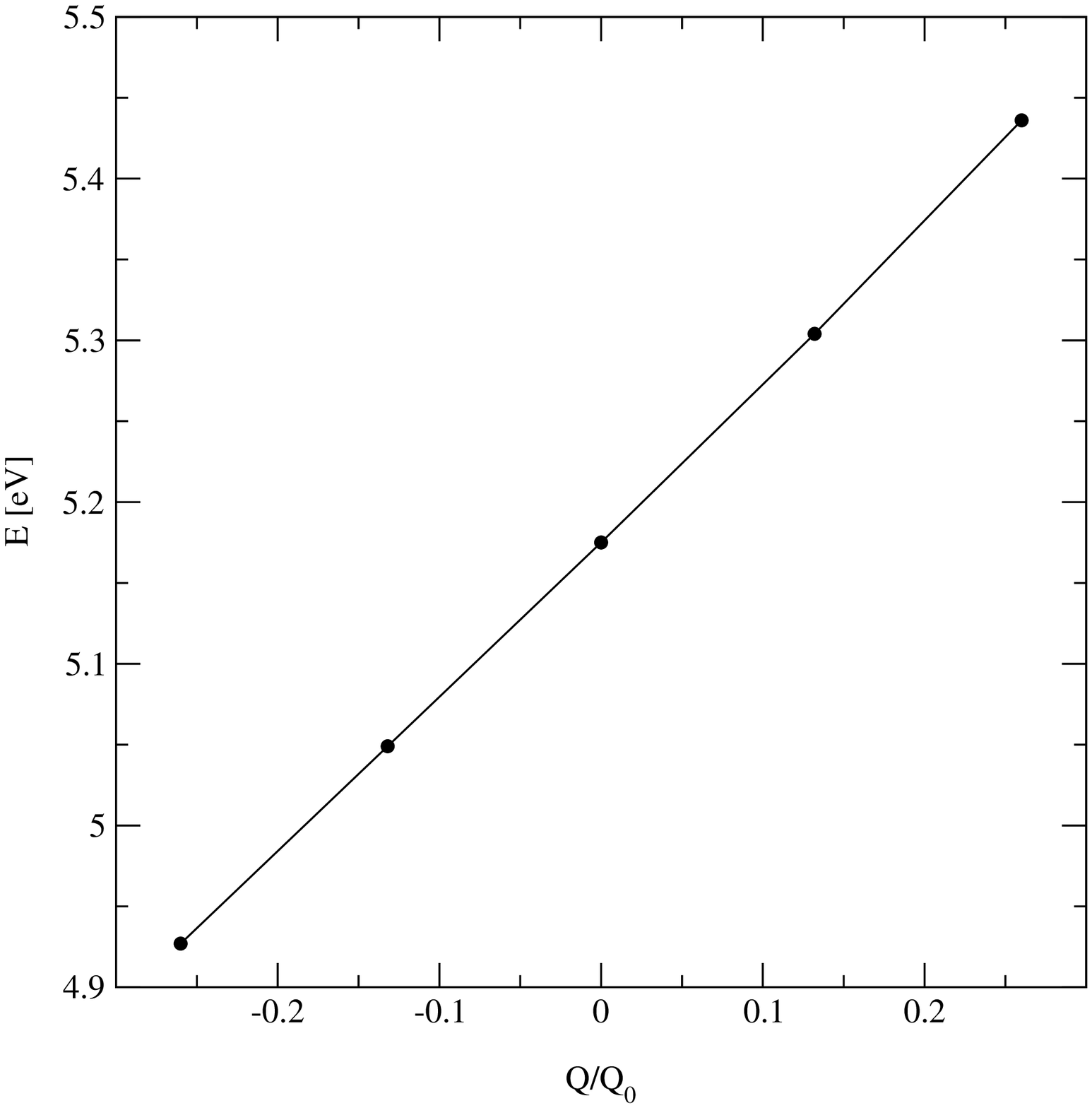,height=7.5cm}
\caption{Dependence of the $^1B_{1u} \leftarrow ^1A_{1g}$
transition energy on the vibrational coordinate for the mode 1.}
\label{2vib}
\end{minipage}
\end{figure}
Next the TDDFT equations are solved in an representation similar to
the RPA equations,
$$
-{\nabla^2 \over 2 m}\phi_i^{\pm} +{\delta {\cal V}\over\delta
n}\phi_i^{\pm}-\epsilon_i\phi_i +{\delta^2 {\cal V}\over \delta n^2}
\delta n \phi_i=
(\epsilon_i \pm\omega)\phi_i^\pm.
$$ 
Here the transition density $\delta n$ and normalization are given by
$$
\delta n = \sum_i\phi_i(\phi_i^++\phi_i^-),
\hbox{~~~}\langle\phi_i^+|\phi_i^+\rangle-\langle\phi_i^-|\phi_i^-\rangle=1.
$$
The equations are solved by the conjugate gradient method for the
generalized eigenvalue problem  \cite{br66}.
In Fig. 4 we show the dependence of transition strengths $f$ and excitation
energies $\omega$ on the coordinates of two of the normal modes.  
We see that the conditions for applying eq. (\ref{f_0k}) and (\ref{P(E)})
are reasonably well satisfied. We may then extract the transition strength
$f_{0k}$ and the width $K_k$ by fitting the $Q_k$-dependence of these
quantities.  The results for the symmetry-allowed vibrations are shown
in shown in Table 2.  

We first discuss the widths.  The empirical values were
obtained by making a three-term Gaussian fit to the absorption data
of ref. \cite{pe68}.  The only vibrations that contribute in lowest order
are the two $A_{1g}$ breathing modes.  The vibrations affect all
three transitions identically; mode 1 has the larger 
amplitude of displacement of the carbon atoms and gives the greater
contribution.  The results agree rather well with the empirical widths.
The magnitude of the widths and its independence of the electronic
state can be understood in very simple terms with the Hueckel model.  
This is to be expected, since the excitation
energy of the electronic states is mainly due to the  
orbital energy difference, and that is describe quite well by the
Hueckel model.  For benzene, the energy difference is related to the hopping
matrix element $\beta$ by 
$
E_{LUMO}-E_{HOMO} =2 \beta.
$.
Allowing changes in the nuclear coordinates, the hopping matrix element
will depend on the distance between neighboring atoms $d$; this may be
parameterized by the form
$$
\beta(d) = \beta_0 \left({d_0\over d}\right)^\alpha.
$$
Then the HOMO-LUMO gap fluctuates due to the breathing mode vibrations with
widths given by
$$
\Delta E = 2\beta_0\alpha {\Delta r\over r}
$$
where $r$ is the radial distance of the carbons from the center and 
$\Delta r$ is at $Q_k=Q_{0k}$ in an $A_{1g}$
mode.  From fitting orbital energies in various conjugated carbon systems
one may extract values $\alpha\approx 2.7$ and $\beta_0 = 2.5$
eV\cite{YB1999}.  
Inserting these
values in the above equation, one obtains 0.145 eV for the widths associated
with mode 1, quite close to the values obtained by TDDFT.  We have included
in the table also the r.m.s.~widths of the Franck-Condon factors obtained by
the CASSCF theory, which gives quite similar results.  One thing 
should be remarked on
the comparison with experiment.  While the theory gives practically
identical widths for all three states, the experimental strength is
significantly narrower for the the $E_{1u}$ excitation, and this seems to
not be understandable in the TDDFT.

Next we examine the transition strengths of the $B$-transitions
induced by the zero-point vibational motion.  In the middle table of 
Table 2 we show the
contributions by the six active vibrational modes.  The main contribution 
for the $B_{2u}$ transition comes from mode
6.  This is also found in the CASSCF theory, and is how the
observed spectrum was interpreted in \cite{hi91}.  
In the case of the $B_{1u}$ excitation, the TDDFT
predicts that the coupling of  mode 8 is dominant.  Experimentally, 
the situation is unclear because the vibrational spectrum of the
excited state is strongly perturbed.  Ref. \cite{hi91} assigns
both mode 6 and mode 8 vibrational involvement.  Irrespective of 
the spectrum of the vibrational modes in the excited state, the total transition
strength is given by the same convolution of the ground state vibrational
wave function.  As in the case of the widths, the induced $B_{1u}$ transition
strength can be understood roughly with the tight-binding model.  The
charge densities are displaced in the vibration, giving the $B_{1u}$
configuration an induced dipole moment just from the atomic geometry.
The Hueckel Hamiltonian of the orbital energy is also affected by the changed separations
between carbons, and that cause a violation of the $B_{1u}$ symmetry.
Finally, the Coulomb interaction, which is mainly responsible for
the splitting of the three electronic states, is affected by the changed
separations.  Of these three mechanisms, only the effect of the 
symmetry-violation in the Hueckel Hamiltonian is important, and mode 8 
crries the largest flutuation in $d$.  Taking the 
same $d$-dependence as before, the strength obtained in the tight-binding
model is 0.05, rather close to the TDDFT result.  The tight-binding model
cannot be used to estimate the very weak $B_{2u}$ transition because the
charge density on the atoms is identically zero.

The lower table gives the empirical transition strengths \cite{pa78}
and comparison
to theory.  The agreement between theory and experiment is quite good
for all states.
For the
weakest transition, the $B_{1u}$, 
the TDDFT gives a transition strength 25\% higher than the empirical
For the case of the $B_{1u}$ transition, the TDDFT 
prediction is within 35\% of the measured 
value.  We also show the previously reported value for the $E_{1u}$
which is within 20\%.  We consider this remarkable success of 
the TDDFT considering that the strengths that range 
over three orders of magnitude.   

In conclusion, we have shown that the TDDFT gives a semiquantitative account of 
the effect of zero-point vibrational motion on the optical absorption
spectrum in benzene. In this respect this extends the possible domain
of utility from the region of infrared absorption, where it is known
that the TDDFT gives a description of transition strengths accurate
to a factor of two or so\cite{be95}.  We are encouraged by these results to
apply the TDDFT to other problems involving the electron-vibrational
coupling.  Perhaps it should be mentioned that not all excitation properties
are reproduced so well in the TDDFT.  In particular, one can not
expect accurate numbers for 
HOMO-LUMO gap of insulators \cite{he99} and the optical rotatory
power of chiral molecules \cite{ya99}.  Of course, there may be better
energy functionals for studying particular properties, and it might
be interesting to examine theories including gradient terms in the
functional.  

We acknowledge stimulating discussions with G. Roepke.  This work was supported
by the Department of Energy under Grant DE-FG06-90ER40561.

\begin{table}[t]
\begin{center}
\begin{tabular}{|l|l|l||l|l|l|}\hline
{\rule[-3mm]{0mm}{8mm} mode}&species&$\bar{\nu}_{\rm obs}\;[{\rm cm}^{-1}]$
&mode&species&$\bar{\nu}_{\rm obs}\;[{\rm cm}^{-1}]$\\ \hline
1&$A_{1g}$&993.1&9&$E_{2g}$&1177.8\\
2&$A_{1g}$&3073.9 &18&$E_{1u}$&1038.3\\
3&$A_{2g}$&1350&19&$E_{1u}$&1484.0\\
12&$B_{1u}$&1010&20&$E_{1u}$&3064.4\\
13&$B_{1u}$&3057&11&$A_{2u}$&674.0\\
14&$B_{2u}$&1309.4&4&$B_{2g}$&707\\
15&$B_{2u}$&1149.7&5&$B_{2g}$&990\\
6&$E_{2g}$&608.1&10&$E_{1g}$&847.1\\
7&$E_{2g}$&3056.7&16&$E_{2u}$&398\\
8&$E_{2g}$&1601.0&17&$E_{2u}$&967\\ \hline
\end{tabular}
\end{center}
\caption{The 20 normal modes of vibration of benzene, numbered according
to  
Wilson \protect\cite{WilsonBook}, with their
symmetry and observed frequency. The data of this table are taken from
\cite{Goodman} where other references can also be found.}
\label{tab_modes}
\end{table}

\begin{table}
\begin{center}
\begin{tabular}{|c|ccc|c|c|}
\hline\hline
Width & \multicolumn{3}{c|} {$A_{1g}$ vibrations }&CASSCF& Exp.\\
$K_k$ (ev)& 1 & 2 & Tot. & &\\
\hline
$^1B_{2u}$ &0.12  & 0.03& 0.15 & 0.14&0.18\\
$^1B_{1u}$&0.12  & 0.03& 0.15 & 0.14 &0.17 \\
$^1E_{1u}$& 0.12     &0.03 &0.15&    & 0.125 \\
\hline
\end{tabular}
\end{center}
\begin{center}
\begin{tabular}{|c|cc|cccc|c|}
\hline\hline
$f_{0k}/10^{-3}$ & \multicolumn{2}{c|}{$B_{2g}$ vib.}&\multicolumn{4}{c|}{$E_{2g}$
vibrations} &\\
TDDFT & 4 & 5 &6 & 7 & 8 & 9 & Total\\
\hline
$^1B_{2u}$&- &- &1.4& 0.2 &-&-&1.6\\
$^1B_{1u}$&- &1.6 &0.4&- &44.&13.&59\\
\hline
\end{tabular}
\end{center}
\begin{center}
\begin{tabular}{|c|cc|c|}
\hline\hline
$f/10^{-3}$&TDDFT&CASSCF&Exp.\\
\hline
$^1B_{2u}$&1.6 &0.5&1.3 \\
$^1B_{1u}$&59 &75 & 90\\
$^1E_{1u}$&1100& & 900-950\\
\hline
\end{tabular}
\end{center}
\caption{Vibrational coupling properties in benzene molecule.
The upper table shows the predicted 
r.m.s.~widths associated with
the breathing mode vibrations.  The total is compared to  the 
CASSCF calculation of ref. \protect\cite{be00}
and to experiment (see text).
In the
middle table, the predicted transition strength associated with the
various vibrations are given, with blank entries having values smaller
than $10^{-4}$.  In the lower table, the predicted total transition
strength is compared with the CASSCF theory and to experiment
\protect\cite{pa78}.
}
\end{table}

\end{document}